\documentclass[fp,twocolumn]{jpsj3}
\usepackage{txfonts}

\usepackage{float}
\usepackage[dvipsnames]{xcolor}

\title{Ground-State Phase Diagram of the Kitaev-Heisenberg Model on a Three-dimensional Hyperhoneycomb Lattice}

\author{Kiyu Fukui\thanks{k.fukui@aion.t.u-tokyo.ac.jp}, Yasuyuki Kato, and Yukitoshi Motome}
\inst{Department of Applied Physics, The University of Tokyo, Bunkyo, Tokyo 113-8656, Japan} 

\abst{
The Kitaev model, which hosts a quantum spin liquid (QSL) in the ground state, was originally defined on a two-dimensional honeycomb lattice, but can be straightforwardly extended to any tricoordinate lattices in any spatial dimensions.
In particular, the three-dimensional (3D) extensions are of interest as a realization of 3D QSLs, and some materials like $\beta$-Li$_{2}$IrO$_{3}$, $\gamma$-Li$_2$IrO$_3$, and $\beta$-ZnIrO$_{3}$ were proposed for the candidates.
However, the phase diagrams of the models for those candidates have not been fully elucidated, mainly due to the limitation of numerical methods for 3D frustrated quantum spin systems. Here we study the Kitaev-Heisenberg model defined on a 3D hyperhoneycomb lattice, by using the pseudofermion functional renormalization group method. We show that the ground-state phase diagram contains the QSL phases in the vicinities of the pristine ferromagnetic and antiferromagnetic Kitaev models, in addition to four magnetically ordered phases, similar to the two-dimensional honeycomb case. Our results respect the four-sublattice symmetry inherent in the model, which was violated in the previous study.
Moreover, we also show how the phase diagram changes with the anisotropy in the interactions. The results provide a reference for the search of the hyperhoneycomb Kitaev materials.
}

\begin{document}
\maketitle

\section{Introduction}\label{sec:intro}
The quantum spin liquid (QSL), which is a quantum disordered state in magnets with fascinating features such as quantum entanglement and fractional excitations, has been studied intensively from both theoretical and experimental points of view~\cite{Diep, Balents2010, Lacroix, Zhou2017}. Despite the long history of research, well-established examples of the QSL are limited, and the realization of the QSL in most of the candidate models and materials is still under debate. The celebrated Kitaev model has brought a revolution to this situation~\cite{Kitaev2006}.  
Despite strong frustration arising from the bond-dependent anisotropic interactions on a two-dimensional (2D) honeycomb lattice, the model is exactly solvable, and the ground state is proven to be a QSL with fractional excitations of itinerant Majorana fermions and localized $Z_{2}$ fluxes; thus, it provides a rare example of exact QSLs in more than one dimension. Moreover, since the feasibility of the model was proposed for spin-orbit coupled Mott insulators~\cite{Jackeli2009}, a number of intensive searches for the candidate materials have been carried out from both theoretical and experimental perspectives~\cite{rau2016, Winter2017, Takagi2019, Motome2020a, Motome2020, Trebst2022}, for instance, for Na$_2$IrO$_3$~\cite{Chaloupka2010, Singh2010, Singh2012, Comin2012, Chaloupka2013, Foyevtsova2013, Sohn2013, Katukuri2014, Yamaji2014, HwanChun2015, Winter2016}, $\alpha$-Li$_2$IrO$_3$~\cite{Singh2012, Chaloupka2013, Winter2016}, and $\alpha$-RuCl$_3$~\cite{Plumb2014, Kubota2015, Winter2016, Yadav2016, Sinn2016}.  
In recent years, a number of new candidates have been proposed, such as cobalt compounds~\cite{Liu2018, Sano2018, Liu2020, Kim2022}, iridium ilmenites~\cite{Haraguchi2018, Haraguchi2020, Jang2021}, and $f$-electron compounds~\cite{Jang2019, Xing2020, Jang2020, Ramanathan2021, Daum2021}. 

While the Kitaev model was originally introduced on the 2D honeycomb lattice, it can be extended to any tricoordinate lattices in any spatial dimensions in a straightforward manner, and in all cases the ground state is an exact QSL. A representative is an extension to a three-dimensional (3D) hyperhoneycomb lattice with space group $Fddd$, shown in Fig.~\ref{fig:lattice}(a), which belongs to a series of extensions of the 2D honeycomb lattice to 3D, dubbed the harmonic honeycomb lattices~\cite{Modic2014}. 
Although the ground state of the 3D hyperhoneycomb Kitaev model is an exact QSL apparently similar to the 2D honeycomb case~\cite{Mandal2009}, finite-temperature properties are qualitatively different: One of two crossovers found in the 2D cases, which is associated with the $Z_2$ fluxes~\cite{Nasu2015}, is replaced by a finite-temperature phase transition in the 3D cases~\cite{Nasu2014, Nasu2014a, Yoshitake2017a, Eschmann2020a, Jahromi2021}. 
This is caused by proliferation of the $Z_2$ fluxes whose excitations form closed loops under the local constraints between the fluxes sharing edges on the 3D lattice~\cite{Kimchi2014a}. 
Similar finite-temperature phase transitions were also found for other 3D extensions of the Kitaev model~\cite{Mishchenko2017, Kato2017, Mishchenko2020, Eschmann2020a}.

On the materials side, $\beta$-Li$_{2}$IrO$_{3}$, where the edge-sharing IrO$_{6}$ octahedra form the 3D hyperhoneycomb network, was initially synthesized and has been investigated intensively as a candidate for the 3D Kitaev QSL~\cite{Takayama2015}. The dominant Kitaev-type interactions in this material was confirmed by the first-principles calculations~\cite{Kim2015, Katukuri2016}.
Although the compound shows a phase transition to a magnetically ordered phase at about 40~K~\cite{Takayama2015, Biffin2014}, the order can be suppressed by the application of a magnetic field~\cite{Ruiz2017} and pressure~\cite{Veiga2017, Majumder2018, Yadav2018}.
Recently, a new candidate $\beta$-ZnIrO$_{3}$ was discovered, and is attracting interests because it does not show any sign of magnetic phase transitions down to 2~K at zero magnetic field and ambient pressure~\cite{Haraguchi2022}. Furthermore, an $f$-electron compound $\beta$-Na$_{2}$PrO$_{3}$, which was synthesized in the past in a different context~\cite{Wolf1988}, was theoretically proposed as a candidate with antiferromagnetic Kitaev interactions~\cite{Jang2020}. 

For understanding the fundamental properties of such 3D hyperhoneycomb Kitaev magnets, there have been a lot of theoretical efforts on some extensions of the Kitaev model~\cite{Lee2014, Lee2014a, Kimchi2014a, Lee2015, Kimchi2015, Lee2016, Huang2018, Ducatman2018, Jahromi2019, Kruger2020, Li2020c, Li2020d}. 
For instance, the Kitaev model with an additional Heisenberg interaction, dubbed the Kitaev-Heisenberg model, was studied by the Luttinger-Tisza method for classical spins~\cite{Lee2014} and by the graph projected entangled-pair states (gPEPS) method for quantum spins~\cite{Jahromi2019}. 
However, comprehensive understanding of the phase diagram and the stability of the QSLs in the 3D cases is still lacking, mainly due to the limited number of efficient theoretical methods for 3D frustrated quantum spin systems.

In this paper, we present our numerical results on the ground state of the Kitaev-Heisenberg model on the hyperhoneycomb lattice obtained by the pseudofermion functional renormalization group (PFFRG) method~\cite{Reuther2009, Reuther2010}. The PFFRG is a powerful numerical method which enables us to perform large-scale calculations for frustrated quantum spin models in any spatial dimensions. Examining the instabilities toward magnetically ordered states by calculating the spin susceptibility, we elucidate the ground-state phase diagram for both isotropic and anisotropic models. For the isotropic case, we find QSL phases around the two pristine Kitaev cases without the Heisenberg interactions, in addition to the four magnetically ordered phases, the ferromagnetic (FM), N\'eel antiferromagnetic (AFM), zigzag AFM, and stripy AFM phases. 
The results look similar to the 2D honeycomb case, but differ from the previous study by the gPEPS method for the 3D hyperhoneycomb case~\cite{Jahromi2019}. We confirm that our results respect the four-sublattice symmetry inherent in the model~\cite{Khaliullin2005, Chaloupka2010, Chaloupka2013, Chaloupka2015, Lee2014}, which was violated in the previous result. Meanwhile, by introducing 
the anisotropy in the interactions, we show that the QSL region is reduced and replaced by the other magnetically ordered state, similar to the previous results obtained by the density matrix renormalization group (DMRG) for the 2D honeycomb case~\cite{Sela2014}.

The structure of this paper is as follows.
In Sect.~\ref{sec:model}, we introduce the Kitaev-Heisenberg model on the hyperhoneycomb lattice and briefly review the previous studies. In Sect.~\ref{sec:method}, we present the essence of the PFFRG method and the calculation conditions. 
We show our results for the ground-state phase diagram for the isotropic case in Sect.~\ref{sec:isotropic} and the anisotropic case in Sect.~\ref{sec:anisotropy}. Finally, Sect.~\ref{sec:summary} is devoted to the summary and perspectives.

\section{Model}\label{sec:model}
\begin{figure}
\includegraphics[width=\columnwidth,clip]{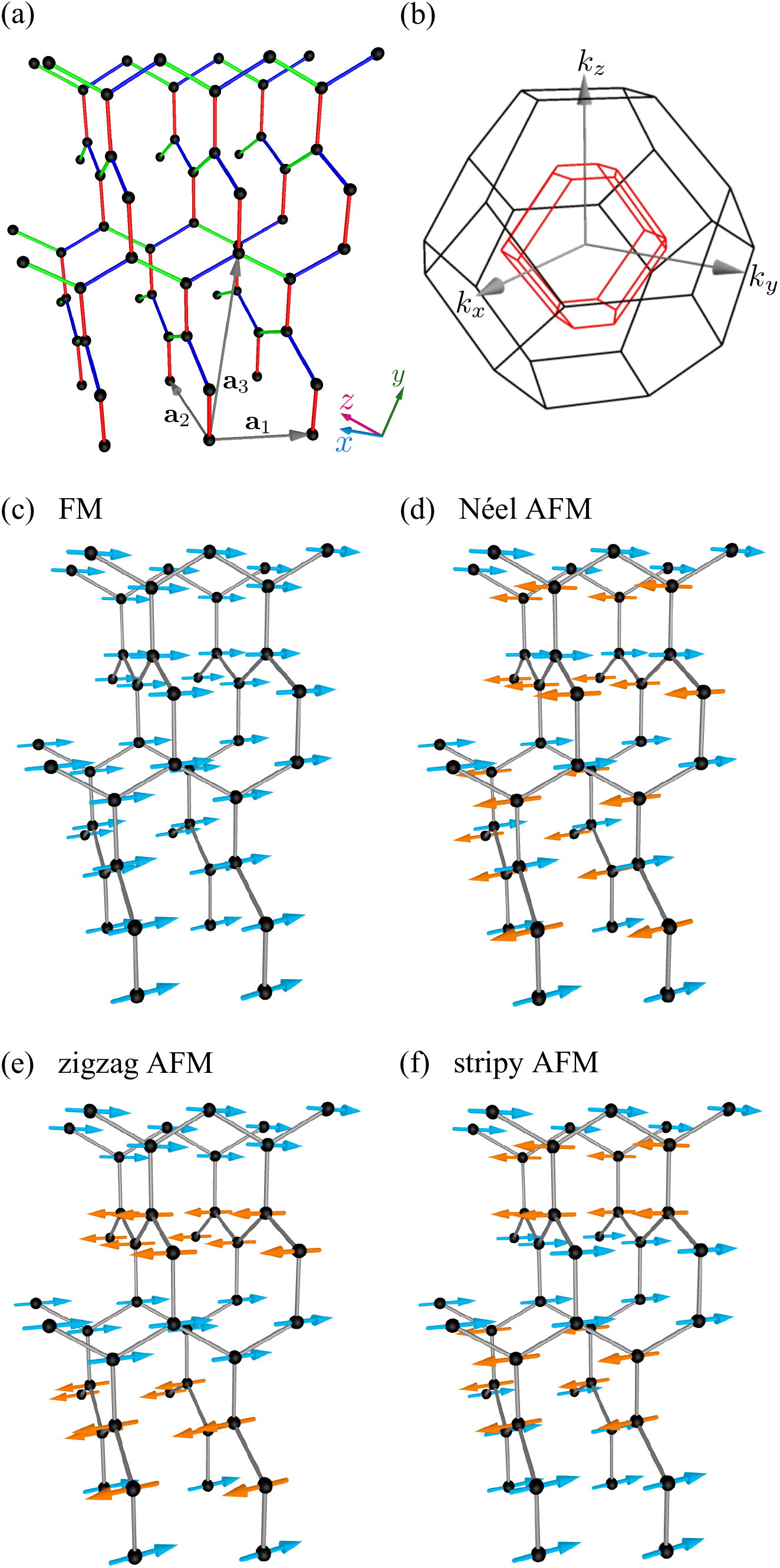}
\caption{(Color online) (a) Schematic picture of the 3D hyperhoneycomb lattice. The blue, green, and red bonds represent the $\mu=x$, $y$, and $z$ bonds, respectively, in the Kitaev-Heisenberg model in Eq.~\eqref{eq:model}. The gray arrows represent the primitive lattice vectors $\mathbf{a}_{1}=(-1/\sqrt{2},\ 1/\sqrt{2},\ -\sqrt{2})$, $\mathbf{a}_{2}=(-1/\sqrt{2},\ 1/\sqrt{2},\ \sqrt{2})$, and $\mathbf{a}_{3}=(\sqrt{2},\ 2\sqrt{2},\ 0)$ in the $xyz$ coordinate shown in the inset.
(b) Brillouin zones for the hyperhoneycomb lattice. The inner red polygon indicates the first Brillouin zone, while the outer black one indicates the Brillouin zone up to the twelfth one. 
(c)--(f) Spin configurations for four magnetically ordered states appearing in the Kitaev-Heisenberg model.
}
\label{fig:lattice}
\end{figure}

We study the Kitaev-Heisenberg model defined on the hyperhoneycomb lattice as a minimal model for the hyperhoneycomb candidate materials. The Hamiltonian is given by 
\begin{equation}
\mathcal{H}=\sum_{\mu=x, y, z}\sum_{\langle i,j\rangle_{\mu}} J_{\mu} \left [2\sin\varphi\ S_{i}^{\mu}S_{j}^{\mu}+\cos\varphi\ \mathbf{S}_{i}\cdot\mathbf{S}_{j}\right ],
\label{eq:model}
\end{equation}
where the summation of $\langle i,j\rangle_{\mu}$ runs over pairs of nearest-neighbor sites $i$ and $j$ connected by $\mu$ bond, and $S_{i}^{\mu}$ is the $\mu$ component of the $S=1/2$ spin operator at site $i$: $\mathbf{S}_{i}=(S_{i}^{x},\ S_{i}^{y},\ S_{i}^{z})$. The first and second terms in Eq.~\eqref{eq:model} represent the Kitaev and Heisenberg interactions, respectively; the ratio of these interactions is parametrized by $\varphi\in[0, 2\pi]$, and the overall strength is given by $J_\mu$. A schematic picture of the model is shown in Fig.~\ref{fig:lattice}(a), in which the $x$, $y$, and $z$ bonds are represented by blue, green, and red, respectively. 
Note that the $z$ bond is crystallographically inequivalent to the rest two on the 3D hyperhoneycomb lattice, while the $x$ and $y$ bonds are related to each other by $C_{2}$ symmetry around the $z$ bonds.

In Eq.~\eqref{eq:model}, the Kitaev interaction is FM for $1<\varphi/\pi<2$, while it is AFM for $0<\varphi/\pi<1$. Meanwhile, the Heisenberg interaction is FM for $1/2<\varphi/\pi<3/2$, while AFM for $0\leq\varphi/\pi<1/2$ and $3/2<\varphi/\pi\leq 2$. There are four special values of $\varphi$: $\varphi/\pi=0$, $1/2$, $1$, and $3/2$. When $\varphi/\pi=1/2$ and $3/2$, the Heisenberg interaction vanishes and the Hamiltonian describes the pristine AFM and FM Kitaev models, respectively, whose ground states are QSLs~\cite{Mandal2009}. Meanwhile, when $\varphi/\pi=0$ and $1$, the Kitaev interaction vanishes and the Hamiltonian corresponds to the AFM and FM Heisenberg models, respectively. 
In these cases, the system has the SU($2$) symmetry, and the FM and N\'eel AFM orders are realized in the ground state. In addition, due to the four-sublattice symmetry~\cite{Khaliullin2005, Chaloupka2010, Chaloupka2013, Chaloupka2015, Lee2014}, there are two more hidden SU($2$) points at $\varphi/\pi=3/4$ and $7/4$ corresponding to $\varphi/\pi=0$ and $1$, respectively.

The Kitaev-Heisenberg model was firstly introduced as an effective model defined on a 2D honeycomb lattice for the 2D candidate materials such as Na$_{2}$IrO$_{3}$ and $\alpha$-Li$_{2}$IrO$_{3}$, and its phase diagram was calculated by using a variety of methods: the exact diagonalization (ED)~\cite{Chaloupka2010,Chaloupka2013,Sela2014}, the DMRG~\cite{Jiang2011,Sela2014}, the slave-particle mean-field approximation~\cite{Schaffer2012}, the tensor network method~\cite{Iregui2014}, the cluster mean-field approximation~\cite{Gotfryd2017}, the high-temperature expansion~\cite{Singh2017}, the quantum Monte Carlo method~\cite{Sato2021}, and the PFFRG method~\cite{Reuther2011a, Fukui2022a}. 
These previous studies showed that the ground-state phase diagram contains four magnetically ordered phases, FM, N\'eel AFM, zigzag AFM, and stripy AFM phases, in addition to the QSL phases in the narrow regions around the two Kitaev points at $\varphi/\pi=1/2$ and $3/2$ where the ground states are the exact QSLs~\cite{Khaliullin2005, Chaloupka2010, Chaloupka2013, Chaloupka2015, Lee2014}.

The model defined on a 3D hyperhoneycomb lattice was introduced, motivated by the synthesis of the hyperhoneycomb candidate $\beta$-Li$_{2}$IrO$_{3}$~\cite{Lee2014}. For the model with classical spins, the ground-state phase diagram was calculated by the Luttinger-Tisza method, and found to contain the four magnetically ordered phases, similar to the 2D honeycomb case~\cite{Lee2014}. Meanwhile, for the quantum spin $S=1/2$ case, the gPEPS calculation showed that the phase diagram contains the QSL phases in the vicinity of the two Kitaev cases in addition to the four magnetically ordered phases~\cite{Jahromi2019}. However, the QSL phases appear in slightly different regions compared to the 2D case: In the 2D honeycomb model, the QSL phases extend from each Kitaev point to both sides of the FM and AFM Heisenberg interactions~\cite{Chaloupka2013, Iregui2014, Gotfryd2017, Sato2021, Fukui2022a}, but they are found almost only on one of the two sides in the gPEPS result for the 3D hyperhoneycomb model.
It should be noted that the four-sublattice symmetry appears to be violated in the gPEPS result, suggesting that the accuracy is insufficient.

\section{Method}\label{sec:method}
In this study, we try to elucidate the ground-state phase diagram of the $S=1/2$ hyperhoneycomb model in Eq.~\eqref{eq:model} by using the PFFRG method.
The PFFRG provides a powerful numerical method for frustrated quantum spin systems~\cite{Reuther2009,Reuther2010}, and has been applied to various models with the Heisenberg~\cite{Reuther2009, Reuther2010}, $XXZ$~\cite{Gottel2012, Buessen2018}, Kitaev-like~\cite{Reuther2011a, Reuther2012, Reuther2014a, Revelli2019a, Fukui2022a}, off-diagonal~\cite{Hering2017, Buessen2019}, long-range dipolar~\cite{Keles2018, Keles2018a, Fukui2022}, and SU($2$)$\times$SU($2$) interactions~\cite{Kiese2019, Gresista2022}. It was applied to 2D systems in the early stage, but later its usefulness was proved for 3D systems~\cite{Iqbal2016, Buessen2016, Buessen2018, Iqbal2018, Iqbal2019, Revelli2019a, Ghosh2019, Niggemann2019, Chillal2020, Zivkovic2021, Hering2022, Noculak2022, Hagymasi2022}. Here we adopt the zero-temperature PFFRG method at the level of one-loop approximation and calculate the $\mu$ component of the static spin susceptibility at momentum $\mathbf{k}$, defined by
\begin{equation}
\chi^{\mu\mu, \Lambda}(\mathbf{k})=\frac{1}{N}\sum_{i,  j}\mathrm{e}^{-\mathrm{i}\mathbf{k}\cdot(\mathbf{r}_{i}-\mathbf{r}_{j})}\chi^{\mu\mu, \Lambda}_{ij}, 
\end{equation}
where $N$ is the number of spins, $\mathbf{r}_{i}$ and $\mathbf{r}_{j}$ represent the real-space coordinates of sites $i$ and $j$, respectively; $\chi^{\mu\mu, \Lambda}_{ij}$ is the spin susceptibility in real space calculated by the PFFRG method as 
\begin{equation}
\chi^{\mu\mu, \Lambda}_{ij}=\int^{\infty}_0\mathrm{d}\tau\ \langle T_{\tau}S^\mu_i(\tau)S^\mu_j(0)\rangle_{\Lambda},
\end{equation}
where $S^{\mu}_{i}(\tau)=\mathrm{e}^{\tau\mathcal{H}}S^{\mu}_{i}\mathrm{e}^{-\tau\mathcal{H}}$, 
$\langle T_{\tau}\cdots\rangle_{\Lambda}$ means the expectation value of the imaginary-time-ordered operators, and $\Lambda$ is the energy cutoff scale in the PFFRG calculations.  
Due to $C_{2}$ symmetry along the $z$ bonds, the $x$ and $y$ components of the susceptibility are related with each other: $\chi^{yy, \Lambda}(\mathbf{k})$ is obtained from $\chi^{xx, \Lambda}(\mathbf{k})$ by interchanging $k_{x}\leftrightarrow k_{y}$ and $k_{z}\leftrightarrow -k_{z}$ simultaneously. 
A magnetic instability is signaled by divergence of $\chi^{\mu\mu, \Lambda}(\mathbf{k})$ at a momentum corresponding to the ordering vector, and the critical value of $\Lambda$ is called the critical cutoff scale $\Lambda_{\mathrm{c}}$. In practice, however, due to the finite system size and the finite frequency grid, the $\Lambda$ dependence of $\chi^{\mu\mu, \Lambda}(\mathbf{k})$ shows a kink or cusp rather than the divergence. Thus, we use such an anomaly to identify the magnetic instability and estimate $\Lambda_{\mathrm{c}}$. Meanwhile, the absence of such an anomaly down to $\Lambda\to 0$ suggests that the system realizes a QSL in the ground state, without showing any magnetic instability. 
We will discuss the relation between $\Lambda_{\mathrm{c}}$ and the transition temperature $T_{\mathrm{c}}$ in Sect.~\ref{sec:isotropic}. 
See Ref.~\citen{Fukui2022} for more details of the PFFRG method we employ. 

In the following calculations, we use the logarithmic frequency grid with 64 positive frequency points between $10^{-4}$ and $250$. We also generate the logarithmic $\Lambda$ grid starting from $\Lambda_{\mathrm{max}}=500$ to $\Lambda_{\mathrm{min}}\simeq10^{-2}$ by multiplying a factor of 0.95 sequentially. In the following calculations for the isotropic case in Sect.~\ref{sec:isotropic}, we include two-particle vertex functions between two sites up to 12th neighbors, which corresponds to a finite-size cluster containing 1034 lattice sites. For the anisotropic case in Sect.~\ref{sec:anisotropy}, we omit the contributions beyond 8th neighbors, which corresponds to a 319-site cluster.

\section{Result}\label{sec:result}
\subsection{Isotropic case}\label{sec:isotropic}
\begin{figure}
\includegraphics[width=\columnwidth,clip]{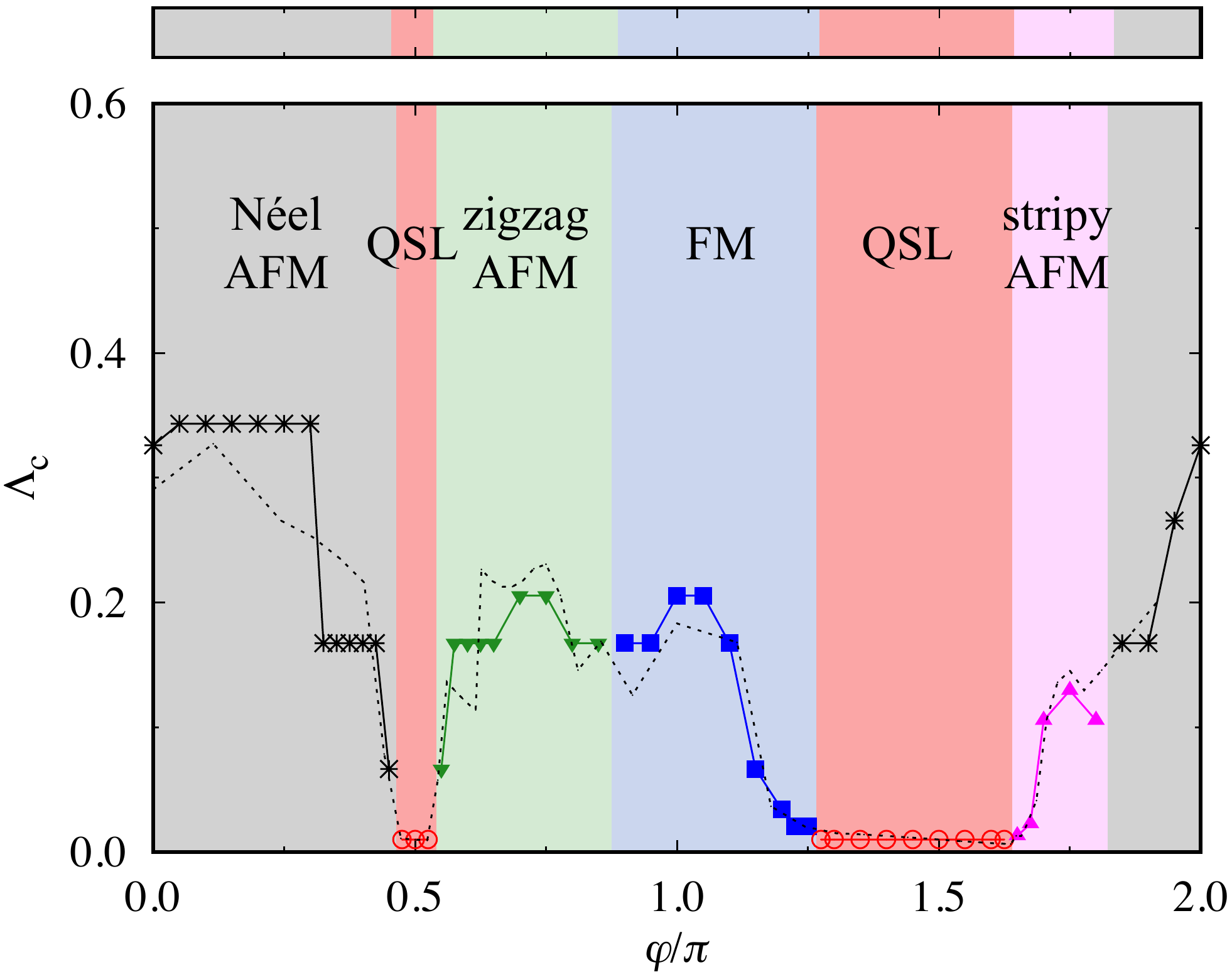}
\caption{(Color online) Ground-state phase diagram and critical cutoff scale $\Lambda_{\mathrm{c}}$ as functions of $\varphi$ for the isotropic Kitaev-Heisenberg model with $J_{x}=J_{y}=J_{z}=1$ defined on the hyperhoneycomb lattice. The black dashed line indicates $\Lambda_{\mathrm{c}}$ obtained by applying the four-sublattice transformation to our results; see Eq.~\eqref{eq:Lambda_trsf}. The top strip shows the ground-state phase diagram for the 2D honeycomb case obtained by the PFFRG method in Ref.~\citen{Fukui2022a}.
}
\label{fig:iso_diagram}
\end{figure}

\begin{figure}[h]
\includegraphics[width=0.9\columnwidth,clip]{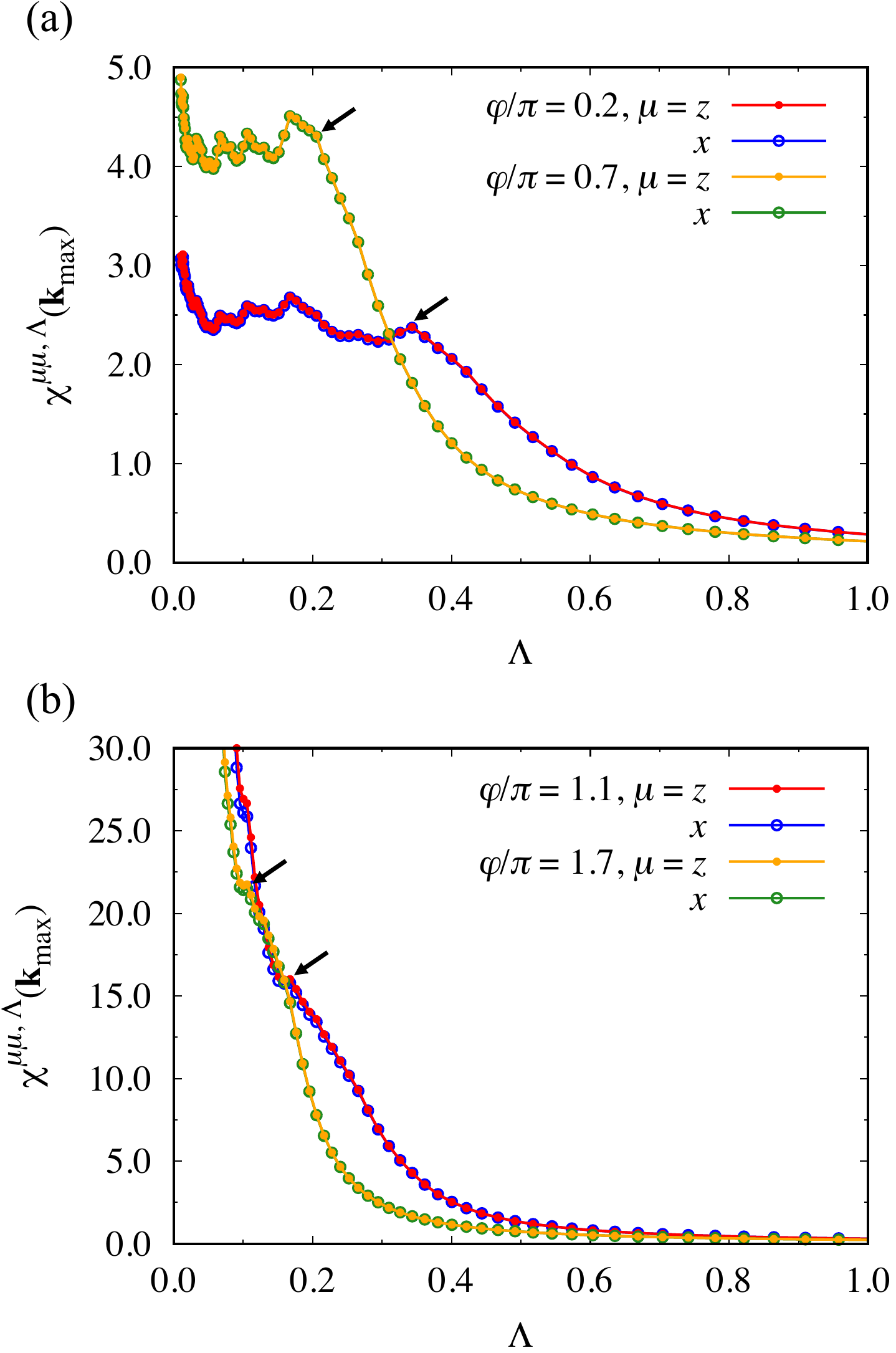}
\caption{(Color online) $\Lambda$ dependences of $\chi^{zz, \Lambda}(\mathbf{k}_{\mathrm{max}})$ and $\chi^{xx, \Lambda}(\mathbf{k}_{\mathrm{max}})$ for the isotropic Kitaev-Heisenberg model at (a) $\varphi/\pi=0.2$ (N\'eel) and $\varphi/\pi=0.7$ (zigzag AFM), and (b) $\varphi/\pi=1.1$ (FM) and $\varphi/\pi=1.7$ (stripy AFM). The black arrows indicate the critical cutoff scale $\Lambda_{\mathrm{c}}$.
}
\label{fig:ordered_flow}
\end{figure}

\begin{figure*}[h!t]
\includegraphics[width=\linewidth,clip]{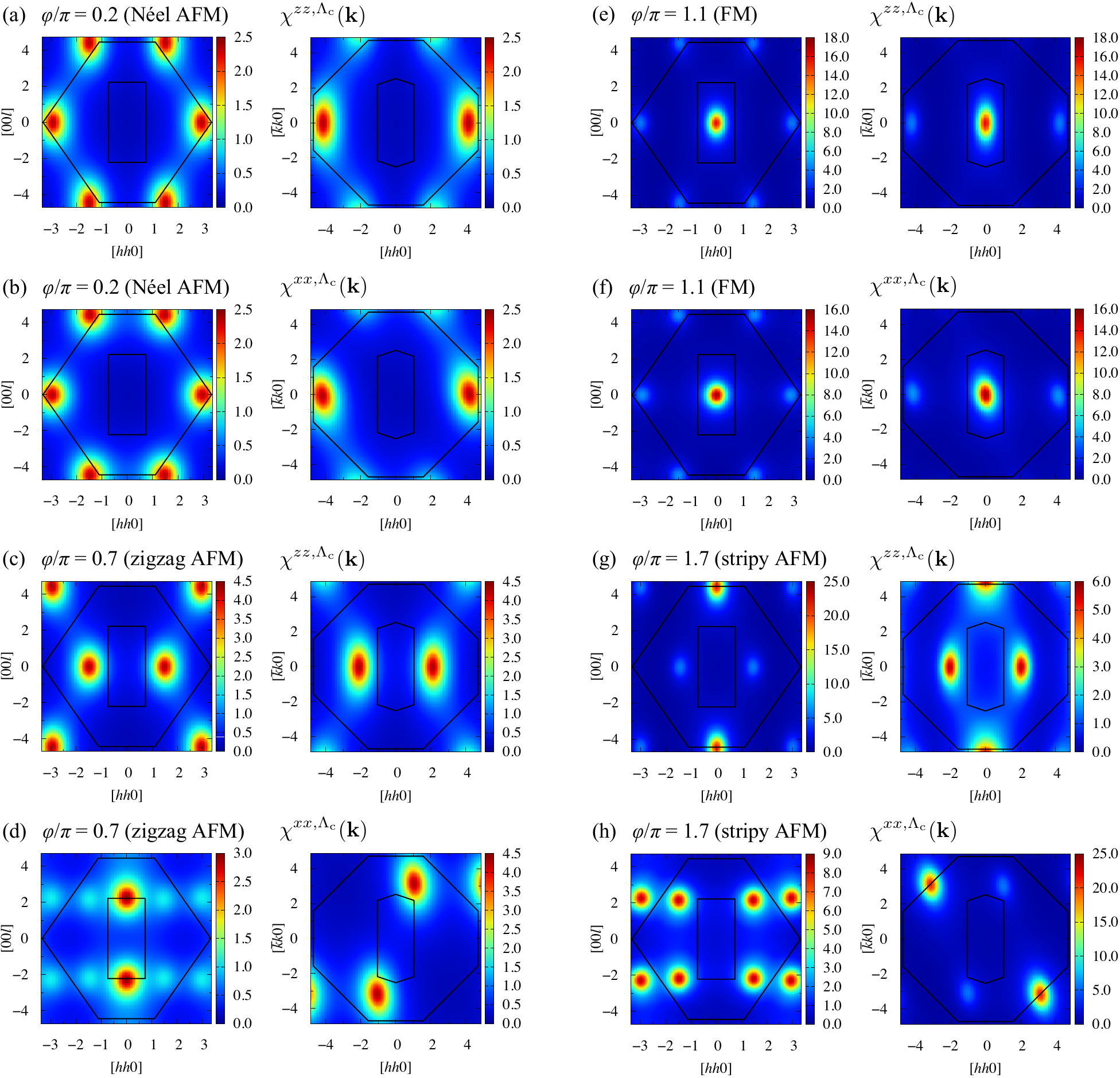}
\caption{(Color online) $\mathbf{k}$ dependences of $\chi^{\mu\mu,\Lambda_{\mathrm{c}}}(\mathbf{k})$ for the isotropic Kitaev-Heisenberg model at (a) and (b) $\varphi/\pi=0.2$ (N\'eel), (c) and (d) $\varphi/\pi=0.7$ (zigzag AFM), (e) and (f) $\varphi/\pi=1.1$ (FM), and (g) and (h) $\varphi/\pi=1.7$ (stripy AFM). (a), (c), (e), and (g) are for $\chi^{zz, \Lambda_{\mathrm{c}}}(\mathbf{k})$, and (b), (d), (f), and (h) are for $\chi^{xx, \Lambda_{\mathrm{c}}}(\mathbf{k})$. 
The left and right panels are the data plotted on the [$hhl$] and [$hk0$] planes, respectively. The inner rectangle and hexagon in the left and right panels, respectively, indicate the first Brillouin zone, while the outer hexagon and octagon in the left and right panels, respectively, indicate the zone up to twelfth one; see Fig.~\ref{fig:lattice}(b). 
}
\label{fig:ordered_kmap}
\end{figure*}

Figure~\ref{fig:iso_diagram} shows the ground-state phase diagram for the isotropic case ($J_{x}=J_{y}=J_{z}=1$) of the hyperhoneycomb Kitaev-Heisenberg model obtained by the PFFRG calculations.  
The phase diagram is determined by the magnetic instabilities appearing in the maximum value of the spin susceptibility, $\chi^{\mu\mu,\Lambda}(\mathbf{k} = \mathbf{k}_{\mathrm{max}})$, where $\mathbf{k}_{\mathrm{max}}$ corresponds to the magnetic ordering vector. 
We find four magnetically ordered phases, N\'eel AFM, zigzag AFM, FM, and stripy AFM phases, as shown in Fig.~\ref{fig:iso_diagram}. The schematic pictures of the spin configurations of these states are shown in Figs.~\ref{fig:lattice}(c)--\ref{fig:lattice}(f), and the locations of $\mathbf{k}_{\mathrm{max}}$ are listed in Table~\ref{table:kmax}.
In Fig.~\ref{fig:iso_diagram}, we plot the values of the critical cutoff energy scale $\Lambda_{\mathrm{c}}$, where $\chi^{\mu\mu,\Lambda}(\mathbf{k}_{\mathrm{max}})$ shows an anomaly. The typical $\Lambda$ dependences in each region are shown in Fig.~\ref{fig:ordered_flow}.
We find that, despite the crystallographic inequivalence between the $z$ and $x$ bonds on the hyperhoneycomb lattice mentioned in Sec.~\ref{sec:model}, $\chi^{zz, \Lambda}(\mathbf{k}_{\mathrm{max}})$ and $\chi^{xx, \Lambda}(\mathbf{k}_{\mathrm{max}})$ show almost the same $\Lambda$ dependences with anomalies at the same $\Lambda_{\mathrm{c}}$ indicated by the black arrows in Fig.~\ref{fig:ordered_flow}. 
Figure~\ref{fig:ordered_kmap} presents the $\mathbf{k}$ dependences of $\chi^{zz, \Lambda}(\mathbf{k})$ and $\chi^{xx, \Lambda}(\mathbf{k})$ at $\Lambda_{\mathrm{c}}$ for the same values of $\varphi$ as in Fig.~\ref{fig:ordered_flow}, which show distinct peaks at $\mathbf{k}_{\mathrm{max}}$, corresponding to each magnetic ordering.
\begin{table}[h]
\caption{
The locations of $\mathbf{k}_{\mathrm{max}}$ in each magnetically ordered phase for the isotropic Kitaev-Heisenberg model. $\mathbf{k}_{\mathrm{max}}$ is the wave vector at which the spin susceptibilities become maximum in the reciprocal space.
}
\label{table:kmax}
\centering
{\renewcommand\arraystretch{1.6}
\begin{tabular}{lcc}
\hline\hline
phase & $\mathbf{k}_{\mathrm{max}}$ for $\chi^{zz, \Lambda}(\mathbf{k})$ & $\mathbf{k}_{\mathrm{max}}$ for $\chi^{xx, \Lambda}(\mathbf{k})$ \\
\hline
N\'eel AFM & $\left(\pm\frac{2\sqrt{2}}{3}\pi, \pm\frac{2\sqrt{2}}{3}\pi, 0\right)$ & $\left(\pm\frac{2\sqrt{2}}{3}\pi, \pm\frac{2\sqrt{2}}{3}\pi, 0\right)$ \\
zigzag AFM & $\left(\pm\frac{\sqrt{2}}{3}\pi, \pm\frac{\sqrt{2}}{3}\pi, 0\right)$ & $\left(\pm\frac{\sqrt{2}}{3}\pi, \mp\frac{2\sqrt{2}}{3}\pi, 0\right)$ \\
FM & $(0, 0, 0)$ & $(0, 0, 0)$ \\
stripy AFM & $(0, 0, \pm\sqrt{2}\pi)$ & $(\pm\sqrt{2}\pi, 0, 0)$ \\
\hline\hline
\end{tabular}
}
\end{table}

\begin{figure}[h]
\includegraphics[width=0.9\columnwidth,clip]{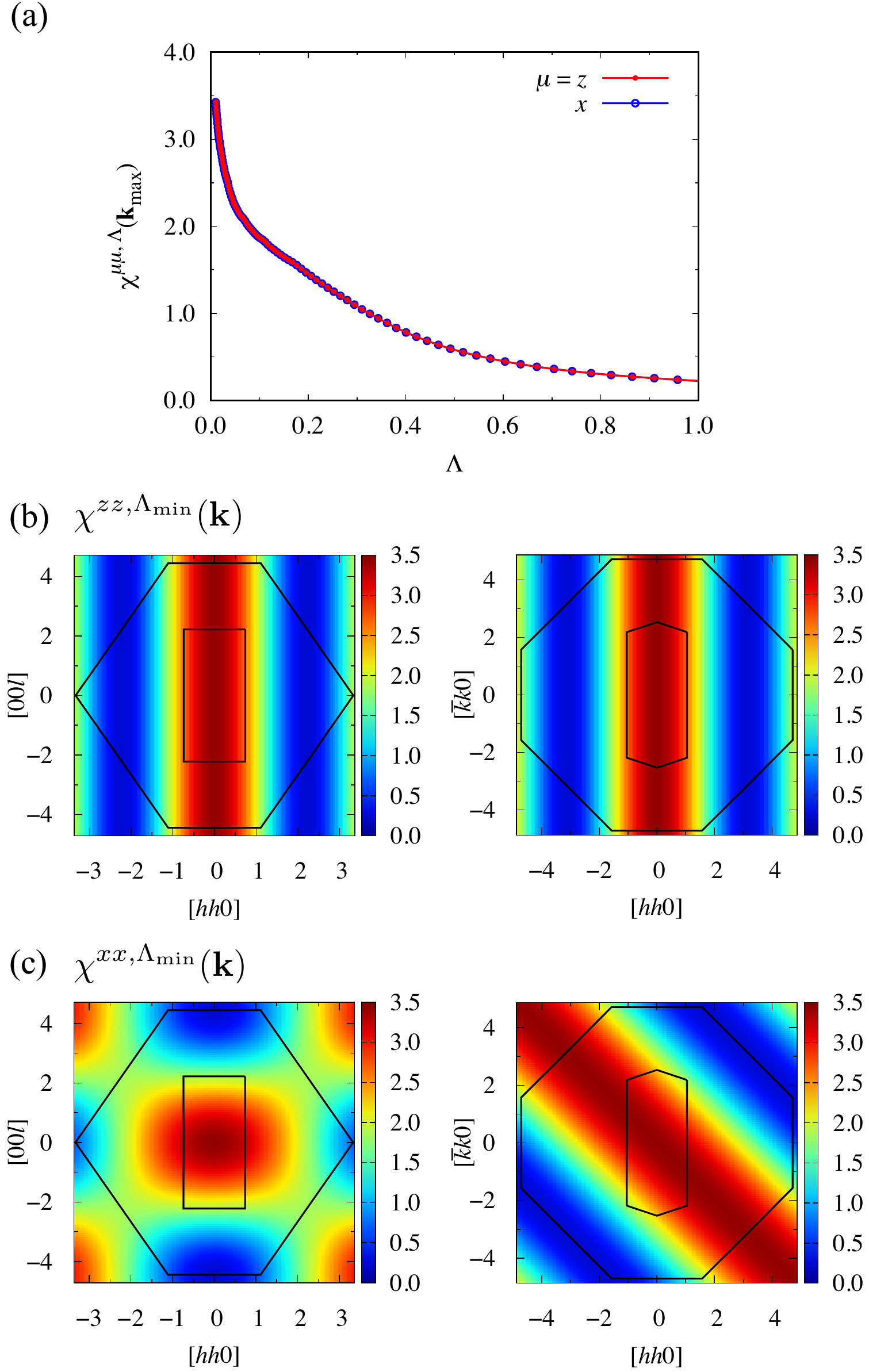}
\caption{(Color online) (a) Spin susceptibilities $\chi^{zz, \Lambda}(\mathbf{k}_{\mathrm{max}})$ and  $\chi^{xx, \Lambda}(\mathbf{k}_{\mathrm{max}})$ as functions of $\Lambda$ for the isotropic FM Kitaev model ($\varphi/\pi=1.5$).  
$\mathbf{k}$ dependences are plotted in (b) for $\chi^{zz, \Lambda}(\mathbf{k})$ and (c) for $\chi^{xx, \Lambda}(\mathbf{k})$ at $\Lambda = \Lambda_{\mathrm{min}}$. 
The notations are common to those in Fig.~\ref{fig:ordered_kmap}.
}
\label{fig:Kitaev}
\end{figure}

In addition to the four ordered phases, we find the Kitaev QSL phase in the two regions including the pristine AFM and FM Kitaev cases at $\varphi/\pi=0.5$ and $1.5$, respectively, as 
indicated by red in Fig.~\ref{fig:iso_diagram}.  
Figure~\ref{fig:Kitaev}(a) shows the $\Lambda$ dependences of $\chi^{zz, \Lambda}(\mathbf{k}_{\mathrm{max}})$ and $\chi^{xx, \Lambda}(\mathbf{k}_{\mathrm{max}})$ for the FM Kitaev case. 
For the AFM case, we obtain the same result since the FM and AFM Kitaev models are equivalent under the gauge transformation~\cite{Kitaev2006}. 
Both $\chi^{zz, \Lambda}(\mathbf{k}_{\mathrm{max}})$ and $\chi^{xx, \Lambda}(\mathbf{k}_{\mathrm{max}})$ show no obvious anomaly down to $\Lambda_{\mathrm{min}}$, suggesting that the ground state is the QSL in consistent with the exact solution~\cite{Mandal2009}.
Figures~\ref{fig:Kitaev}(b) and \ref{fig:Kitaev}(c) show the $\mathbf{k}$ dependences of $\chi^{zz, \Lambda}(\mathbf{k})$ and $\chi^{xx, \Lambda}(\mathbf{k})$ at $\Lambda_{\mathrm{min}}$, respectively. 
In this FM Kitaev case, $\chi^{zz(xx), \Lambda}(\mathbf{k})$ shows the maximum at $k_{x}+k_{y}=0$ ($-k_{y}+k_{z}=0$) with arbitrary $k_{z}$ ($k_{x}$). 
The results indicate that $\chi^{zz, \Lambda}(\mathbf{k})$ and $\chi^{xx, \Lambda}(\mathbf{k})$ are well approximated by $\propto \cos[(k_{x}+k_{y})/\sqrt{2}]+\mathrm{const}.$ and $\propto \cos[(-k_{y}+k_{z})/\sqrt{2}]+\mathrm{const}.$, respectively. This means that the spin correlations are negligible beyond nearest neighbors, as in the 2D honeycomb case~\cite{Baskaran2007}, which is also consistent with the exact solution~\cite{Mandal2009}. We find similar behaviors even in the presence of small Heisenberg interactions for $0.4625\lesssim\varphi/\pi\lesssim 0.5375$ and $1.2625\lesssim\varphi/\pi\lesssim1.6375$, as shown in Fig.~\ref{fig:iso_diagram}. 
The latter region around the FM Kitaev point is considerably wider than the former around the AFM Kitaev point, as seen in the 2D honeycomb case~\cite{Chaloupka2013, Iregui2014, Gotfryd2017, Sato2021, Fukui2022a}.

We note that the previous PFFRG results for the 2D honeycomb case tend to overestimate the QSL regions compared to the ED or DMRG results~\cite{Reuther2011a, Fukui2022a}, presumably due to differences in the numerical methods and the system sizes. 
The same is likely to be true for the present 3D hyperhoneycomb case. 
In any case, our results show that the Kitaev QSL phases extend from each Kitaev point to both sides of the FM and AFM Heisenberg interactions as in the previous studies for the 2D honeycomb model~\cite{Chaloupka2013, Iregui2014, Gotfryd2017, Sato2021, Fukui2022a}. 
This is qualitatively different from the previous results using the gPEPS method where those extend to almost only one of the two sides~\cite{Jahromi2019}.

For further comparison with the previous gPEPS study, we examine the four-sublattice symmetry which the model in Eq.~\eqref{eq:model} satisfies~\cite{Khaliullin2005, Chaloupka2010, Chaloupka2013, Chaloupka2015, Lee2014}. 
Under the transformation, $\sin\varphi$, $\cos\varphi$,  and $\varphi$ in Eq.~\eqref{eq:model} are transformed as
\begin{align}
(\sin\varphi', \cos\varphi') &= \mathcal{N} (\sin\varphi+\cos\varphi, -\cos\varphi),\\
\varphi'&=\mathrm{arctan}[-\tan\varphi-1],
\end{align}
where $\mathcal{N}$ is the normalization so that $\sin^{2}\varphi'+\cos^{2}\varphi'=1$: $\mathcal{N} = \{(\sin\varphi+\cos\varphi )^2+\cos^{2}\varphi\}^{-1/2}$. Hence, the value of $\Lambda_{\mathrm{c}}$ at $\varphi$ is transformed into $\Lambda_{\mathrm{c}}'$ at $\varphi'$ as
\begin{equation}
\Lambda_{\mathrm{c}}'(\varphi')=\mathcal{N}\Lambda_{\mathrm{c}}(\varphi).
\label{eq:Lambda_trsf}
\end{equation}
The values of $\Lambda_{\mathrm{c}}'(\varphi')$ obtained from our numerical estimates of $\Lambda_{\mathrm{c}}(\varphi)$ are plotted by the dashed line in Fig.~\ref{fig:iso_diagram}. 
We find that $\Lambda_{\mathrm{c}}'(\varphi')$ almost agree with the original $\Lambda_{\mathrm{c}}(\varphi)$.  
We note considerable deviations especially for $0.0 \leq \varphi/\pi \lesssim 0.5
$, but it would be attributed to the finite frequency grid in our PFFRG calculations~\cite{Fukui2022a}.
Thus, our results respect the required four-sublattice symmetry, whereas the previous gPEPS ones which predicted largely asymmetric QSL regions do not.

It is worth noting that the ground-state phase diagram in Fig.~\ref{fig:iso_diagram} is very similar to that of 2D honeycomb case obtained by the PFFRG method~\cite{Fukui2022a}, which is shown in the top strip of the Fig.~\ref{fig:iso_diagram}.  
This is presumably because (i) both 2D honeycomb and 3D hyperhoneycomb lattices are tricoordinated, 
(ii) the spin correlations are nonzero only between the nearest neighbors in the Kitaev limit~\cite{Baskaran2007}, and (iii) all the ordered states induced by the Heisenberg interaction have commensurate ordering vectors.  
Indeed, the classical energies and the ground-state phase diagram obtained by the Luttinger-Tisza method are exactly the same for the two lattices~\cite{Kruger2020}. 
A difference between 2D honeycomb and 3D hyperhoneycomb cases is expected to be pronounced at finite temperature.
For the 2D honeycomb case, the transition temperature is strictly zero in the four SU($2$) cases with $\varphi/\pi=0.0$, $0.75$, $1.0$, and $1.75$ because of the Mermin-Wagner theorem~\cite{Mermin1966}, while it becomes finite for the 3D hyperhoneycomb case. In addition, in the Kitaev QSL regions, a topological transition by loop proliferation of the flux excitations is expected to occur at finite temperature in the 3D case~\cite{Nasu2014, Nasu2014a, Yoshitake2017a, Eschmann2020a, Jahromi2021}, whereas it is absent and only a crossover is left in the 2D case~\cite{Nasu2015}. 

In this respect, it is interesting to note that $\Lambda_{\mathrm{c}}$ can be regarded as an estimate of the transition temperature $T_{\mathrm{c}}$, by assuming a relation between the energy scale $\Lambda$ and temperature $T$ as $T\simeq \frac{\pi}{2} \Lambda$, which holds for large $\Lambda$ and $T$~\cite{Iqbal2016,Buessen2016,Buessen2019}. 
Indeed, the PFFRG results for the 2D honeycomb model with large spin $S=50$ qualitatively reproduced the onset temperature of the quasi-long-range order obtained by classical Monte Carlo simulations~\cite{Fukui2022a}. 
Thus, we may conclude that $\Lambda_{\mathrm{c}}$ in Fig.~\ref{fig:iso_diagram} gives an estimate of $T_{\mathrm{c}}$ for the 3D hyperhoneycomb Kitaev-Heisenberg model. 
Then, an intriguing issue is whether the PFFRG can predict the finite-temperature topological transition expected to occur in the Kitaev QSL regions. 
The value of $T_{\mathrm{c}}$ was estimated at $\sim 0.0078$ in our energy unit~\cite{Nasu2014}, which corresponds to $\Lambda_{\mathrm{c}} \sim 0.005$. This value is smaller than the minimum value of $\Lambda$, $\Lambda_{\mathrm{min}}\simeq10^{-2}$, in Fig.~\ref{fig:Kitaev}. 
We further calculate the susceptibilities down to $\Lambda \sim 0.0014$, but do not find any anomaly. This suggests that the finite-temperature topological transition by loop proliferation leaves no trace in the $\Lambda$ dependence in the present PFFRG calculation. This is presumably a limitation of the correspondence between $T$ and $\Lambda$ in the zero-temperature PFFRG. Alternatively, this might be due to the fact that our one-loop PFFRG method incorporates only up to two-body interactions, wheres the flux is a ten-body quantity in the hyperhoneycomb lattice.

\subsection{Anisotropic case}\label{sec:anisotropy}
Finally, we investigate the effect of anisotropy in the magnetic interactions on the ground-state phase diagram. In  this section, we consider the region $1.5\leq\varphi/\pi\leq2.0$, where the Kitaev and Heisenberg couplings are FM and AFM, respectively, and the effect of anisotropy was studied for the 2D honeycomb model~\cite{Sela2014}. Assuming $J_{x}=J_{y}$, we parametrize the anisotropy as 
\begin{equation}
J_{x}=J_{y}=(3-J_{z})/2,
\label{eq:anisotropy}
\end{equation}
and change $J_{z}\in [0, 3]$; the system becomes disconnected one-dimensional chains of $x$ and $y$ bonds in the limit of $J_z\to 0$, while it becomes independent dimers of $z$ bonds in the limit of $J_z\to 3$.
 
\begin{figure}
\includegraphics[width=\columnwidth,clip]{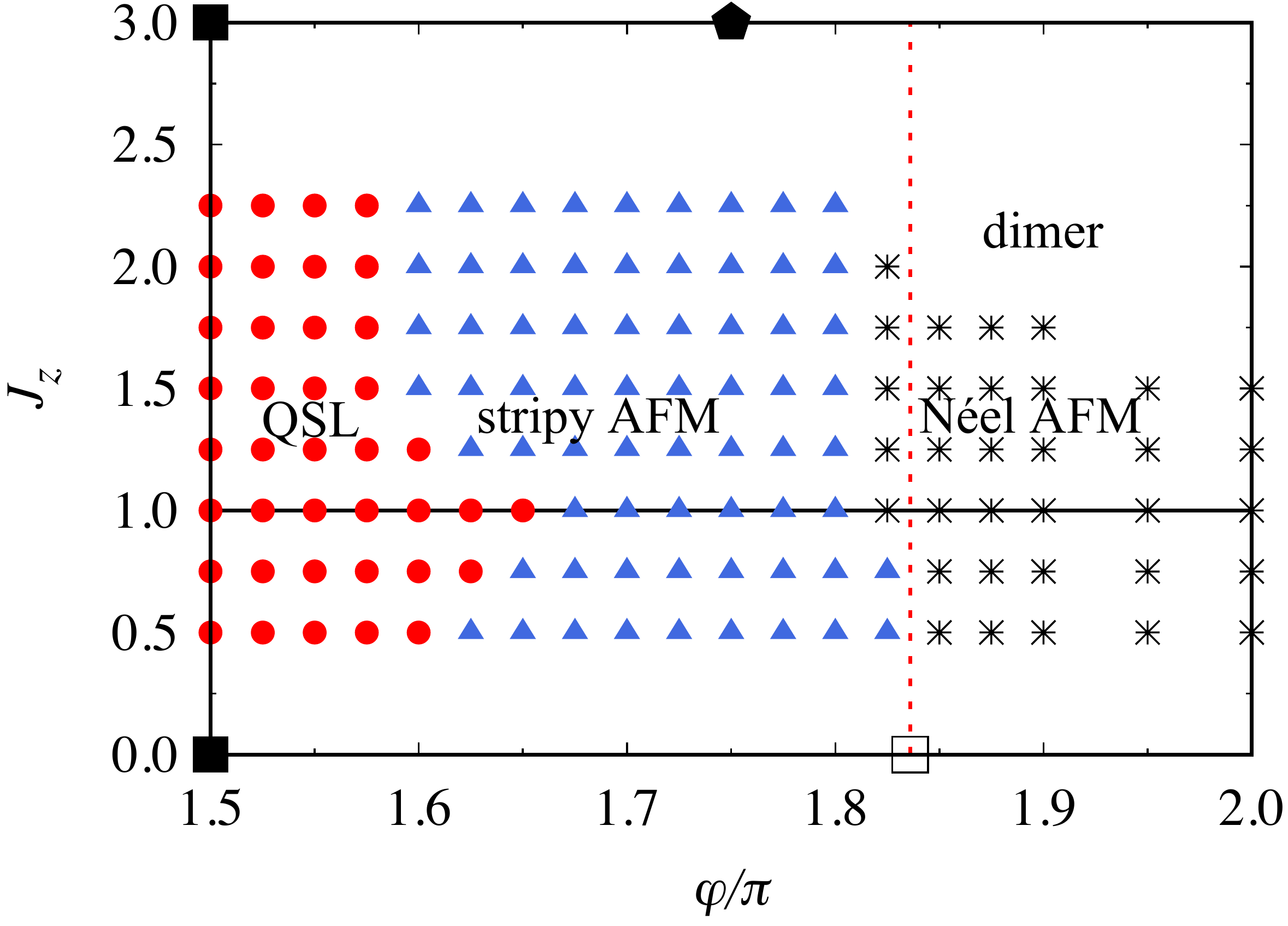}
\caption{(Color online) Ground-state phase diagram of the anisotropic Kitaev-Heisenberg model in Eq.~\eqref{eq:model} with Eq.~\eqref{eq:anisotropy} on the plane of $\varphi/\pi$ and $J_{z}$. 
The symbols on the top and bottom of the phase diagram indicate the phase boundaries expected in the anisotropic limits of $J_z\to 0$ and $J_z\to 3$; see the main text for the details.
The red dashed line indicates the phase boundary obtained by the Luttinger-Tisza method for classical spins.
}
\label{fig:aniso_diagram}
\end{figure}

Figure~\ref{fig:aniso_diagram} shows the ground-state phase diagram obtained by the PFFRG method. The data are limited within the region of $0.5\leq J_z\leq 2.25$ because outside this region the PFFRG method cannot correctly detect anomalies of the susceptibilities due to the small energy scales in the anisotropic cases. 
When introducing the anisotropy, the QSL region is narrowed and replaced by the stripy AFM, while the phase boundary between the stripy and N\'eel AFM is almost intact. 
In the large $J_{z}$ and $\varphi$ region, the N\'eel AFM phase is replaced by the dimer phase, where the spin susceptibility does not show any anomaly except for a broad hump in the $\Lambda$ dependence. 
These results are qualitatively similar to the previous ones for the 2D honeycomb case~\cite{Sela2014}. 

For comparison, we plot the phase boundaries in the anisotropic limits in Fig.~\ref{fig:aniso_diagram}. 
The two filled squares at $(\varphi/\pi, J_z) = (1.5, 0.0)$ and $(1.5, 3.0)$ indicate that the QSL is unstable against infinitesimally small Heisenberg interactions in both limits, as shown for the 2D case~\cite{Sela2014}. 
The phase boundary between the QSL and stripy AFM states in our results should be extrapolated to these two points, despite the lack of data in the anisotropic regions. 
Meanwhile, the filled pentagon at $(\varphi/\pi, J_z) = (1.75, 3.0)$ indicates the phase boundary between the stripy AFM and dimer states in the limit of $J_z\to 3$, which is obtained by comparing the energy of triplet and singlet states for a two-site dimer. 
The open square in the opposite limit of $J_z\to 0$ indicates the phase boundary between the stripy and N\'eel AFM states, numerically estimated  for the 2D case~\cite{Sela2014}. 
This agrees well with the estimate by the Luttinger-Tisza method for classical spins, as indicated by the vertical dashed line in Fig.~\ref{fig:aniso_diagram}. 
The phase boundary in our results appear to be consistent with these two limits.

\section{Summary and perspectives}\label{sec:summary}
To summarize, we have studied the Kitaev-Heisenberg model defined on the 3D hyperhoneycomb lattice by using the PFFRG method. We clarified the ground-state phase diagram for the model with isotropic interactions by changing the ratio between the Kitaev and Heisenberg interactions. We identified the regions of two QSL phases around the two pristine Kitaev cases, in addition to the four magnetically ordered phases, N\'eel AFM, zigzag AFM, FM, and stripy AFM phases. Our results respect the four-sublattice symmetry, and are similar to those of the 2D honeycomb model obtained by the PFFRG method~\cite{Fukui2022a}, in contrast to the previous study by the gPEPS method~\cite{Jahromi2019}. 
We also investigated the effect of the spatial anisotropy and showed that the QSL phase is the most stable for the isotropic case and shrinks when the anisotropy is introduced. These results are also qualitatively similar to those of the 2D honeycomb model~\cite{Sela2014}. 

Our results provide a reference for not only the understanding of the existing candidate materials but also the search and design of the hyperhoneycomb Kitaev materials. 
The candidate material $\beta$-Li$_{2}$IrO$_{3}$ shows an incommensurate noncoplanar magnetic order at low temperature~\cite{Takayama2015, Biffin2014}, which does not appear in our phase diagram in Fig.~\ref{fig:iso_diagram}. This suggests the importance of additional interactions beyond the Kitaev-Heisenberg model. 
Indeed, recent experimental results show that a symmetric off-diagonal interaction, called the $\Gamma$ interaction, is not negligible~\cite{Majumder2019, Ruiz2021, Halloran2022}. In this compound, it was also shown that an external pressure destroys the magnetic order and stabilizes a dimer state~\cite{Veiga2017, Majumder2018}. 
Since the hyperhoneycomb lattice is not crystallographically isotropic, the pressure may enhance or reduce the anisotropy of the interactions, and hence, an extension of our phase diagram in Fig.~\ref{fig:aniso_diagram} could be useful for understanding such a transition. 
An external magnetic field also suppresses the magnetic order~\cite{Ruiz2017}, which urges theoretical studies including the effect of magnetic fields, as intensively discussed for the 2D case. It would also be intriguing to understand the QSL-like state in the new candidate $\beta$-ZnIrO$_3$~\cite{Haraguchi2022}. All these extensions can be handled by the PFFRG method, and left as subjects for future study. 

\section*{Acknowledgments}
The authors thank T. Misawa, J. Nasu, and T. Okubo for fruitful discussions.
Parts of the numerical calculations have been done using the facilities of the Supercomputer Center, the Institute for Solid State Physics, the University of Tokyo, the Information Technology Center, the University of Tokyo, and the Center for Computational Science, University of Tsukuba.
This work was supported by Japan Society for the Promotion of Science (JSPS) KAKENHI Grant Nos. 19H05825 and 20H00122.

\bibliographystyle{jpsj}
\bibliography{library}

\end{document}